\documentclass[%
reprint,
amsmath, amssymb, 
longbibliography,
aps,
]{revtex4-1}

\usepackage{graphicx}
\usepackage{subfigure}
\usepackage{float}
\usepackage{dcolumn}
\usepackage{bm}
\usepackage{color}
\usepackage[english]{babel}
\usepackage[latin1]{inputenc}
\usepackage{lmodern}

\begin{document}

\preprint{APS/123-QED}
\title{Features of the extreme events observed in the all-solid state laser with a saturable absorber}

\author{Carlos R. Bonazzola}


\author{Alejandro A. Hnilo}
\altaffiliation[Corresponding author: ]{ahnilo@citedef.gob.ar}
\author{Marcelo G. Kovalsky}
\affiliation{CEILAP, Centro de Investigaciones en L\'{a}seres y Aplicaciones, UNIDEF (MINDEF-CONICET);\\
 J.B. de La Salle 4397, (1603) Villa Martelli, Argentina. 
}%

\author{Jorge R. Tredicce}

\affiliation{Departamento de F\'{i}sica, Facultad de Ciencias Exactas y Naturales, Universidad de Buenos Aires;\\ Intendente G\"{u}iraldes 2160, Ciudad Aut\'{o}noma de Buenos Aires, Argentina}. \\

\date{\today}
             
\begin{abstract}

Extreme 
events (sometimes also called optical rogue waves), in the form of pulses of extraordinary intensity, are easily observed in its 
chaotic regime if the Fresnel number of the cavity is high. This result suggests that the nonlinear interaction among transverse 
modes is an essential ingredient in the formation of extreme events in this  type of lasers, but there is no theoretical 
description of the phenomenon yet. We report here a set of experimental results on the regularities of these extreme events, to 
provide a basis for the development of such a description. Among  these results, we point out here: i) the decay of the 
correlation across the transversal section of the laser beam, and ii) the appearance of extreme events even if the time elapsed 
since the previous pulse is relatively short (in terms of the average inter-pulse separation), what indicates the existence of 
some unknown mechanism of energy storage. We hypothesize that this mechanism is related with the imperfect depletion of the gain 
by some of the transversal modes. We also present evidence in support of this hypothesis.

\end{abstract}

\begin{description}
\item[PACS numbers]
42.65.Sf, 43.55.Xi, 42.60.Mi, 05.45.Tp 
\end{description}

\pacs{Valid PACS appear here}
\maketitle

\section{INTRODUCTION}\label{sec1}

In recent years there has been a growing interest in extreme events (EEs) in various disciplines \cite{albeverio_extreme_2006}. 
The first reliable measurements of freak or rogue 
waves (i.e. unexpectedly large oceanic waves, as high as 30 m from crest to trough) in the beginning of the 90s, were followed 
by intensive research on what had previously been considered a near-mythological phenomenon. In 2007
Solli et al. introduced the concept of 
optical rogue waves to describe large fluctuations in the edge of the spectrum of the light generated by the propagation of seed 
pulses in a microstructured 
optical fibre, thus posing an analogy between 
the optical pulses and their oceanic 
counterparts \cite{solli_optical_2007}. The authors did so based on, firstly, the L-shaped statistics of the optical events (long 
tailed distributions imply the existence of EEs, that, although rare, are observed with non-negligible probability), 
and, secondly, the theoretical approach they employed to describe the results of the experiment. This description is based on 
the Nonlinear Schr\"{o}dinger Equation (used in this case  to describe the propagation of pulses in the optical fiber). Rogue waves appear after a modulational instability is reached \cite{pelinovski_extreme_2008}. Since then, similar phenomena 
have been found and studied in a wide variety of optical systems, under the broad denomination 
of optical rogue waves or EEs, meaning, roughly speaking, extreme fluctuations in the value of an optical 
field \cite{dudley_instabilities_2014}. However, we 
must warn the Reader that this does not imply \textit{a priori} a common generation 
mechanism of these phenomena, or even a common single definition. In fact, even in Oceanography, where the term \textit{rogue 
wave} was originally coined, a unified definition does not 
yet exist \cite{ruban_rogue_2010}. The task becomes more difficult when considering diverse physical systems beyond the oceans 
and analogies between these phenomena must be dealt with carefully (a thorough discussion on this topic can be found in 
\cite{dudley_instabilities_2014}).
Regardless of their connection with their oceanic counterparts, optical EEs are intrinsically interesting. They have been 
studied in extended optical systems (including a linear and a nonlinear experiment -a laser beam focused into a perturbed 
multimode glass fiber, and an optical cavity that uses a liquid crystal light-valve as nonlinear medium, respectively-) 
\cite{residori_rogue_2012}, in optically injected semiconductor lasers \cite{bonatto_deterministic_2011, 
zamora-munt_rogue_2013,reinoso_extreme_2013}, and in lasers with saturable absorbers, both \textit{fast} (in a Kerr-lens mode 
locked
laser) \cite{kovalsky_extreme_2011} and \textit{slow} (in a passively Q-switched all-solid-state Nd:YAG+Cr:YAG laser) 
\cite{hnilo_extreme_2011}. For a comprehensive review, we refer the Reader to \cite{akhmediev_recent_2013}. In particular, 
we reported the existence of EEs in a system very similar to the latter: a Nd:YVO$_{4}$+Cr:YAG laser 
\cite{bonazzola_optical_2013}, a device of a wide practical interest. We showed that EEs are observed in chaotic regimes 
with high dimension of embedding, high Fresnel number for the cavity and complex spatial transverse patterns of the spot. The 
standard theoretical approach based on rate equations for a single mode \cite{tang_deterministic_2003}, though able to describe 
many of the dynamical 
features of this system, does not predict the existence of EEs. This suggests that 
interaction of transverse modes is a necessary condition for the formation of EEs in this system. One of the aims of this 
contribution is 
to provide further proof for this claim. Another goal is to perform a thorough exploration of the regimes in which EEs appear, 
and of the EEs themselves, to guide the theoretical modelling of the problem, as well as to outline key points that should be 
predicted by such model. With these objectives in mind, we analyse some of the spatiotemporal features of the dynamics of these 
events. The understanding of the mechanism of formation of EEs in this system 
may conceivably lead to its control and, eventually, to useful applications. 

In order to quantitatively define EEs, we adopt the criterion that the pulses with a peak intensity exceeding the 
mean by more than 4 times the standard deviation of the peak intensity distribution (4$\sigma$ threshold) are EEs. This 
criterion is somewhat arbitrary; however, we will show later that it is appropriate not 
only because the events that fulfill it are among the highest in a particular regime of laser operation, but also because they 
exhibit peculiar dynamical features. Besides, we are specifically interested in those EEs appearing in regimes with long tailed 
histograms, where EEs appear more frequently than in Gaussian distributions. In other words, we are interested in distributions with a kurtosis higher than 3.

This paper is organized as follows: in section \ref{sec2} we describe the laser and the methods to record and analyse of the 
time series.  Also, the setup to observe transversal coherence domains (\ref{sec2-1}) and the spatial correlation among 
different sections of the spot (\ref{sec2-2}). In section \ref{sec3}, we discuss the main experimental results, namely: i) the plots of 
intensities in partial sections of the spot show that the EEs are not linked to an unique transverse pattern (\ref{sec312}); ii)  in the 
dynamical regimes with EEs, the two-point time correlation decays to zero at a distance nearly half the size of the spot, while 
in periodic regimes such decay is not observed (\ref{sec311b}); iii) 
the study of the time intervals between successive pulses and the return maps of peak pulse intensity suggest that the EEs occur in a relatively well defined manifold in the phase space and, in consequence, that there is some hope to predict 
them (\ref{sec32}); this claim is supported by the behavior of the intensity return maps (\ref{sec33}); iv) heterodyne interferograms indicate the existence of domains of transversal coherence, confirming the 
results of (i) (\ref{sec34}); v) a summarizing diagram shows that the regimes with EEs are chaotic or hyperchaotic with embedding 
dimension larger than 6, and that they have relatively complex transverse patterns (\ref{sec35}). In what follows, we assume the 
Reader to be familiar with the essentials of self Q-switching theory \cite{siegman_lasers_1986, koechner_solid-state_2006, yariv_quantum_1975}.

\section{Experimental setup}\label{sec2}

\subsection{The laser}\label{sec2a}
\begin{figure}
\includegraphics[scale=0.29]{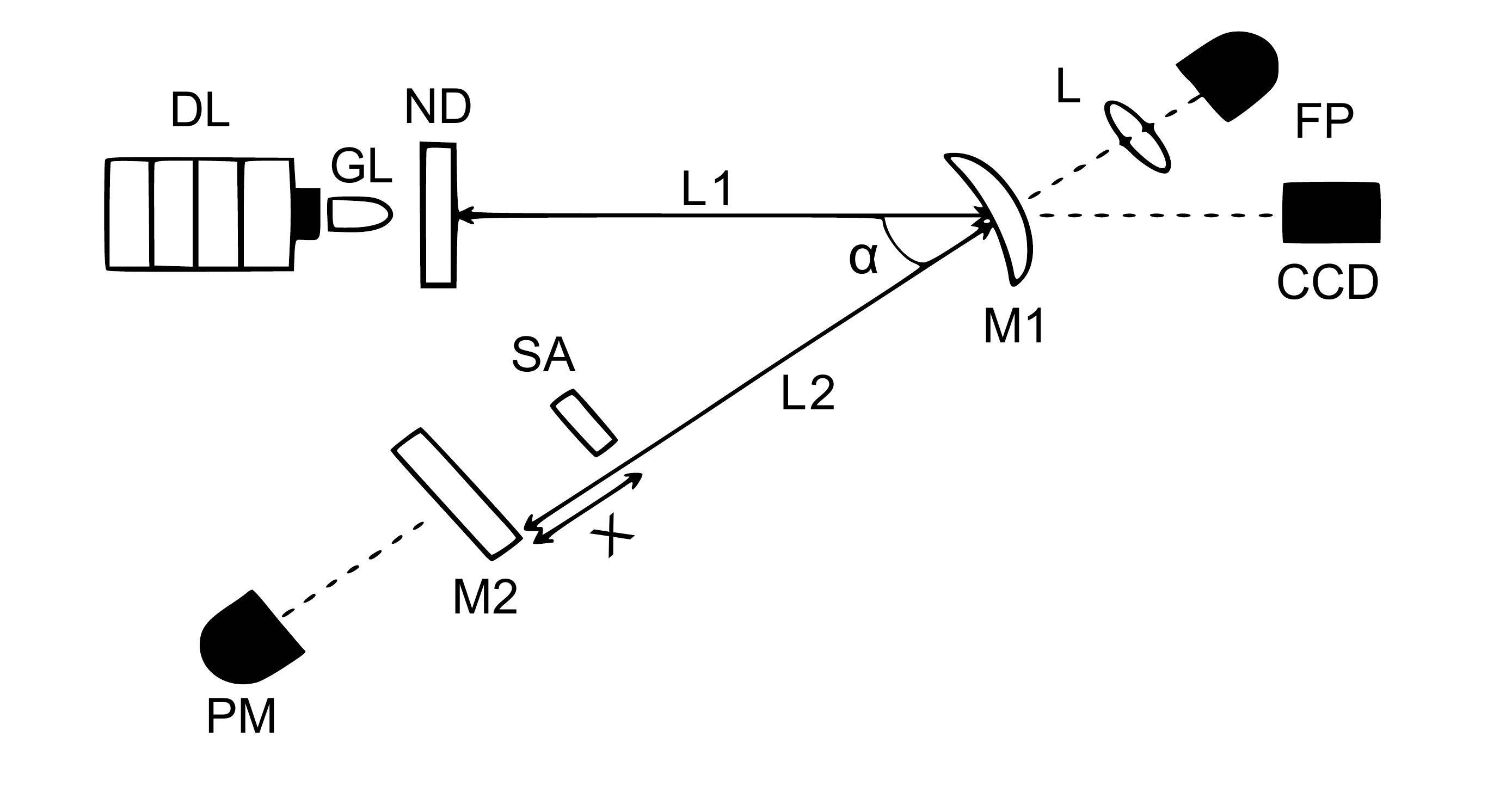}
\caption{\label{fig:laser} Laser setup. LD: pump laser diode,2 W CW @808 nm; GL: GRIN lens; ND: Nd:YVO4 slab (active
medium); M1: folding mirror (R $=$ 250 mm); M2: output mirror (plane); SA: Cr:YAG crystal, transmission (unbleached) 90$\%$; L: focusing lens; FP: fast photodiode; CCD: camera for recording spot images; PM: 
power meter, $\alpha = 20^{\circ}$; L1 $=$ 130 mm; L2$=$75 mm. The position 
X of the SA is variable, to obtain different dynamical regimes.}
\end{figure}

The setup is shown in figure 
\ref{fig:laser}. The output of a 2W (@808 nm ) CW laser diode is collimated by a GRIN lens and focused down to a spot 0.12 mm  
diameter into a Nd:YVO$_{4}$ crystal, 1\% doped and appropriately coated, mounted on a water-
cooled copper heatsink. The V-shaped laser cavity has a folding high reflectivity concave mirror (R$=$1 m) and 
a plane output coupler (reflectivity = 98\%). The mode size varies between the mirrors, with the waist near the 
output coupler. The average output power is measured with the power meter PM placed after the output 
coupler. 
A solid-state saturable absorber (Cr:YAG crystal, 90\% transmission) is placed between the folding mirror and the output coupler 
at a variable distance X. By varying its position, the mode size at the saturable absorber changes and hence the condition of saturation. As the 
absorber is displaced along the arm of the cavity, different dynamical regimes, such as periodic behavior (periods 2, 4, 6), 
a period-three stable window \cite{bonazzola_optical_2013-1} and chaotic regimes with and without EEs 
\cite{bonazzola_optical_2013} are observed. 

One of the output beams at the folding mirror is focused into a pin fast photodiode (100 ps risetime), connected to a PC 
oscilloscope (PicoScope\textregistered  6403B: 500 MHz bandwidth, 5 GS/s, memory of 1 GS). Time series of the self-Q-switching pulse intensities with several thousand pulses are recorded. These series are later analysed with the 
TISEAN software package \cite{hegger_practical_1999,schreiber_surrogate_2000}, in order to calculate the dimension of embedding 
and the Lyapunov exponents. A VGA CCD camera with 60 fps time resolution connected to a PC allows to measure the total intensity distribution. The whole setup is mounted on an optical table 
with interferometric stability. 

\subsection{Interferograms}\label{sec2-1}

\begin{figure}
\includegraphics[scale=0.5]{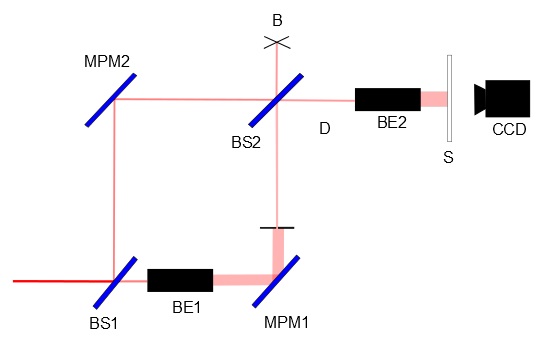}
\caption{\label{fig:MZ} modified Mach-Zehnder interferometer used to study transversal coherence of the spot. BS1: beam splitter, 
transmission 96$\%$; MPM1 and MPM2: metallic plane 
mirrors; BE1 and BE2: 5$\times$ beam expanders; BS2: 50/50 beam splitter; D: diafragm; S: screen; CCD: CCD camera; B: block.}
\end{figure}

In order to study the transversal coherence of the spot in different dynamical regimes we use a modified Mach-Zehnder 
interferometer. The laser beam is collimated with a convergent lens (F=100 mm) placed at 90 mm from the output coupler, and 
then directed into the interferometer shown in fig. \ref{fig:MZ}. The configuration of the Mach-Zehnder is such that it allows 
us to obtain, by means of a 5$\times$ beam expander, a superposition of the spot with a magnified partial section of itself. This output is 
in turn magnified (so as to make easier the task of discerning interference lines), projected on a screen, and registered with 
a CCD camera.

\subsection{Measurement of intensities in partial sections of the spot}\label{sec2-2}

The comparison of intensities in different partial sections of the spot (by means of plots of the intensity in one section vs. 
the intensity in the other) is an interesting way to study its 
transverse dynamics in a pulse to pulse evolution, allowing us to determine, for example, whether EEs are linked to specific 
transverse patterns. A complementary approach to characterize the transverse dynamics is 
the two point spatial correlation. Unlike the plots of intensities in different sections, it provides a magnitude averaged over 
a whole time series, and it is useful to identify the presence of spatial domains and eventually detecting 
spatio-
temporal chaos. 

In order to study these features, the loss beam is magnified with a 5$\times$ beam expander and then directed to
a 50/50 beam-splitter. Both reflected and transmitted beams are then measured with two photodiodes, which are covered by masks 
with 1mm diameter pinholes to limit the measured section, and 
connected to the PC oscilloscope. One of these photodiodes (which we call photodiode A) is fixed, and is kept measuring the 
center of the spot, 
while the other (photodiode B) is mounted on a micro-metric translation stage, so that it can measure the intensity in different 
regions of the spot. The 
integrated intensity of the whole spot is at the same time registered with the photodiode FP. This scheme allows us to measure, 
simultaneously, the total intensity of the spot and the intensity in two specific regions.

\section{RESULTS}\label{sec3}

\subsection{Correlation among different regions of the spot}\label{sec31}

\subsubsection{Intensity plots in different sections of the spot}\label{sec312} 

\begin{figure}
\includegraphics[scale=0.13]{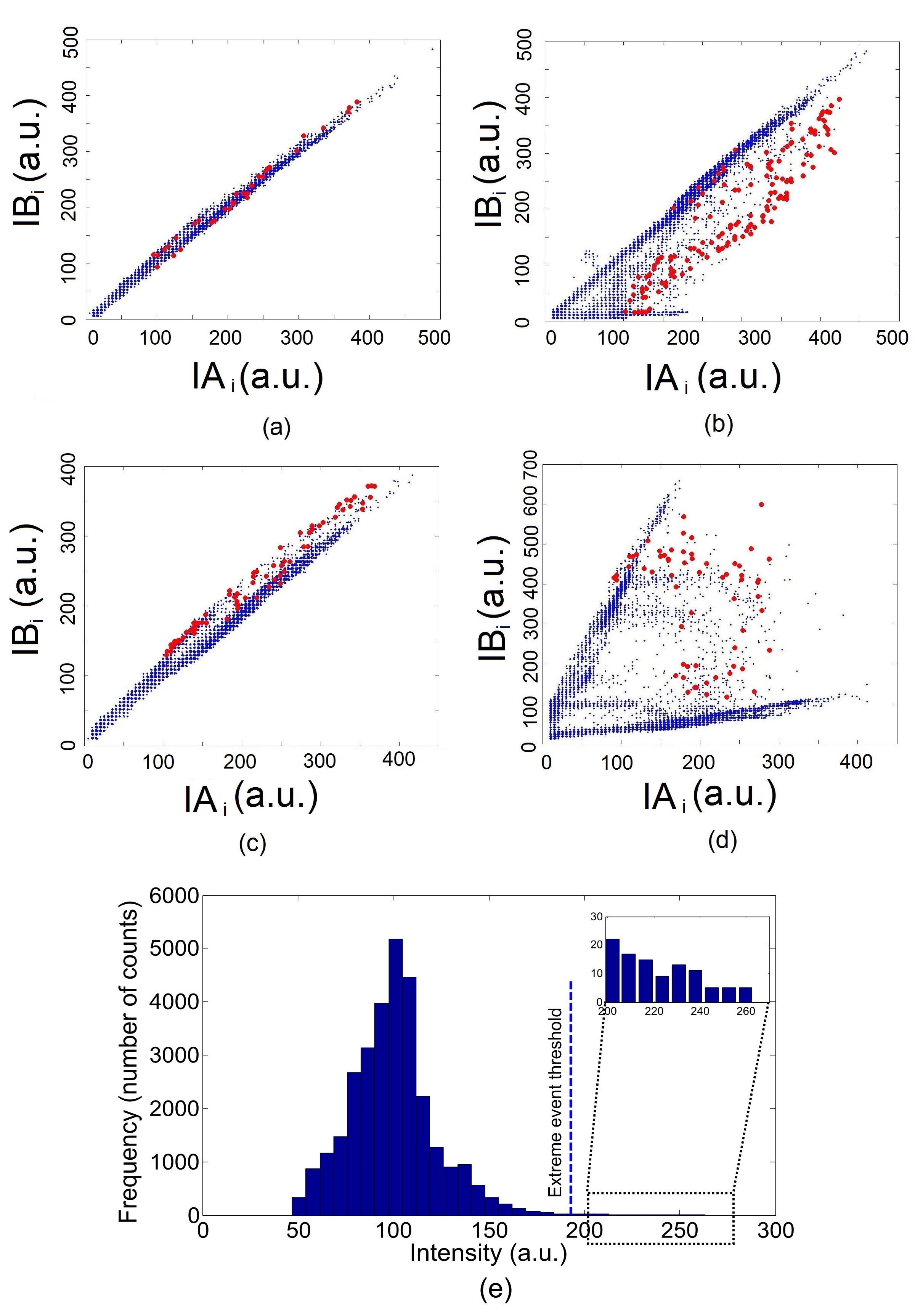}
\caption{\label{fig:secsparcs} Peak pulse intensity measured in the photodiode B (IB$_i$) vs peak pulse intensity in photodiode A 
(IA$_i$) for a chaotic regime with EEs, for different values of 
horizontal separation $d$: (a) $d=0$ mm, (b) $d=1$ mm, (c) $d=3$ mm, (d) $d=8$ mm. Thick 
(red online) dots indicate EEs; (e) histogram of total pulse intensity for the series, the kurtosis of the 
distribution is 6.02. The data belong to the same time series of the regime depicted with blue circles in fig. \ref{fig:corresp}. 
Thus, the corresponding laser spot is the one shown in the left of the lower row of that figure.} 
\end{figure}

We study the evolution of the transverse pattern from pulse to pulse, by analysing pairs of time series of peak intensities in 
different sections of the spot. We call $IA_i$ ($IB_i$) the peak intensity measured by photodiode A (B) for the $i_{th}$ pulse of 
the series. Fig. 
\ref{fig:secsparcs} (a-d) shows $IB_i$ vs. $IA_i$ for different distances $d$ for a regime with EEs, and fig. \ref{fig:secsparcs} 
(e) is a histogram of the total peak intensity. As it is expected, for $d=0$  (i.e., when both photodiodes register the same 
region of the spot), the relation between 
intensities is linear (see fig. \ref{fig:secsparcs} (a)); for $d=1$ mm (fig. \ref{fig:secsparcs} (b)), the plot spreads, showing 
an irregular distribution; for $d = 3$ mm (fig. 
\ref{fig:secsparcs} (c)) it becomes 
practically linear again; and finally, for $d=8$ mm (fig. 
\ref{fig:secsparcs} (d)) it spreads even more than in the previous case. Four main features arise from these plots: i) the 
transverse pattern changes from 
pulse to pulse in an irregular fashion (a behavior like this has been reported for periodic regimes of similar 
systems in \cite{wei_modeling_2010} and \cite{wei_spatial_2004}; ii) some regions of the spot show correlation (e.g. fig. 
\ref{fig:secsparcs} (c)), 
while others are almost uncorrelated (e.g. figures \ref{fig:secsparcs} (b) and (d)); iii) EEs appear in different 
zones of the plots: this implies that the spatial configuration of the spot changes from one EE to another, i.e. the 
EEs are not associated to a 
single transverse pattern. Moreover, they occur in a wide range of values both for IA and IB, i.e.: some EEs 
correspond to a high 
intensity value in section A and low in section B, others to viceversa, and others 
with a low or high intensity intensity in both sections. This means that, in the case that EEs were related to specially 
brilliant regions, these would change position from one EE to the other, and would not be associated to a specific 
location in the transverse pattern; iv) despite the seemingly random distribution in figures (b) and (d), some regularities 
arise 
from the plot. E.g., most of the events are contained within two approximately straight lines: this might be an indication 
of the predominance of two different transverse configurations that appear with a higher frequency than others, but with 
different total intensities each time (i.e. every point in one of the lines would be associated to the same spatial 
configuration as the others in the same line, but with a different total intensity).

\subsubsection{Spatial correlation}\label{sec311b}

We calculate the two-point spatial correlation for pairs of peak pulse intensity time 
series (corresponding to different pairs of sections of the spot). It is defined as:

\begin{equation}
C(IA,IB)=\frac{\displaystyle\sum_{i=1}^{N} (IA_i-\overline{IA})(IB_i-\overline{IB})}{\sqrt{\displaystyle\sum_{i=1}^{N} 
(IA_i-\overline{IA})^2\displaystyle\sum_{i=1}^{N}(IB_i-\overline{IB})^2}}
\label{ec:corresp}
\end{equation}

where $\overline{IA}$ ($\overline{IB}$) is the average intensity over the whole time series IA (IB). 

\begin{figure}
\includegraphics[scale=0.09]{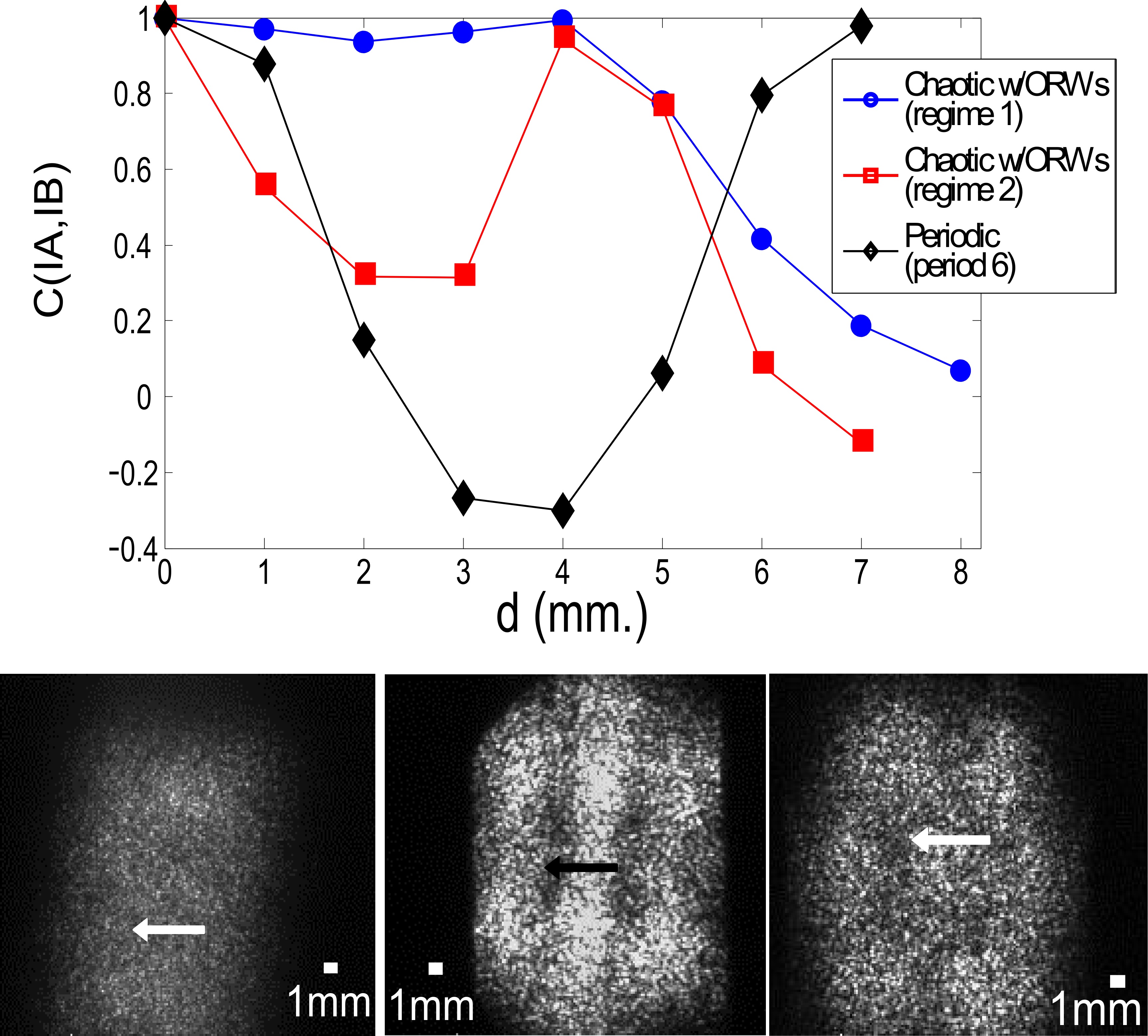}
\caption{\label{fig:corresp} Up: spatial correlation among different sections of the spot as a function of their transversal 
separation $d$ for two different chaotic regimes with EEs (regime 1, depicted by circles) and (regime 2, depicted by squares) and a periodic regime, with period 6 
(diamonds). Down: laser spots for the regimes in the upper plot: from left to right, chaotic regime 1, chaotic regime 2, 
periodic; the arrow indicates the direction in which the different regions of the spot were sampled, its origin 
shows the position of the fixed photodiode.}
\end{figure}

Fig. \ref{fig:corresp} shows the values of $C(IA,IB)$ for two different chaotic regimes with EEs which we call regime 1 (circles) 
and 2 (squares) and a 
periodic regime (diamonds). The different regimes are obtained by adjusting the position of the saturable absorber inside the 
laser cavity, as explained in section \ref{sec2a}. The x-axis 
corresponds to the transversal distance $d$ between the sections observed by both photodiodes. 

The line depicted by circles is typical of the regimes with EEs. The coefficient 
$C(IA,IB)$ remains close to 1 for nearly half of the spot size, and then drops to almost 0 near the edge of the spot. This 
behavior differs from that observed in spatiotemporal chaotic systems with a large number (of the order of tens) of modes 
(which can be roughly estimated in optical systems by the Fresnel number), characterized by a rapid exponential decay of the 
correlation \cite{arecchi_experimental_1990,hohenberg_chaotic_1989}, and/or a narrow peak with a short 
correlation length \cite{huyet_spatiotemporal_1995}. The behavior observed in our laser is consistent with the existence of a 
relatively few transverse modes defining 
domains of correlation that spread through large areas of the spot from pulse to pulse. 

Some of the regimes with EEs show a non monotonous behavior, as the one represented by 
squares in fig. \ref{fig:corresp}: a local peak away from the center, before decaying for larger $d$.  
This does not contradict our previous statement, but implies these regimes are dominated by transverse patterns different from 
those involved in the aforementioned ones, arguably occupying areas with more complex shapes. This is confirmed by the shape of 
the whole spot registered by the CCD camera. 

For comparison, we also show in fig. \ref{fig:corresp} the spatial correlation for the case of a periodic regime with period 6 
(diamonds). The correlation decays at first, but near the edge of the spot it rises again to a value close to 1. In other 
periodic regimes, the spatial correlation remains near 1, except in a few narrow valleys.

\subsection{Temporal intervals between succesive pulses}\label{sec32}

\begin{figure}
\includegraphics[scale=0.23]{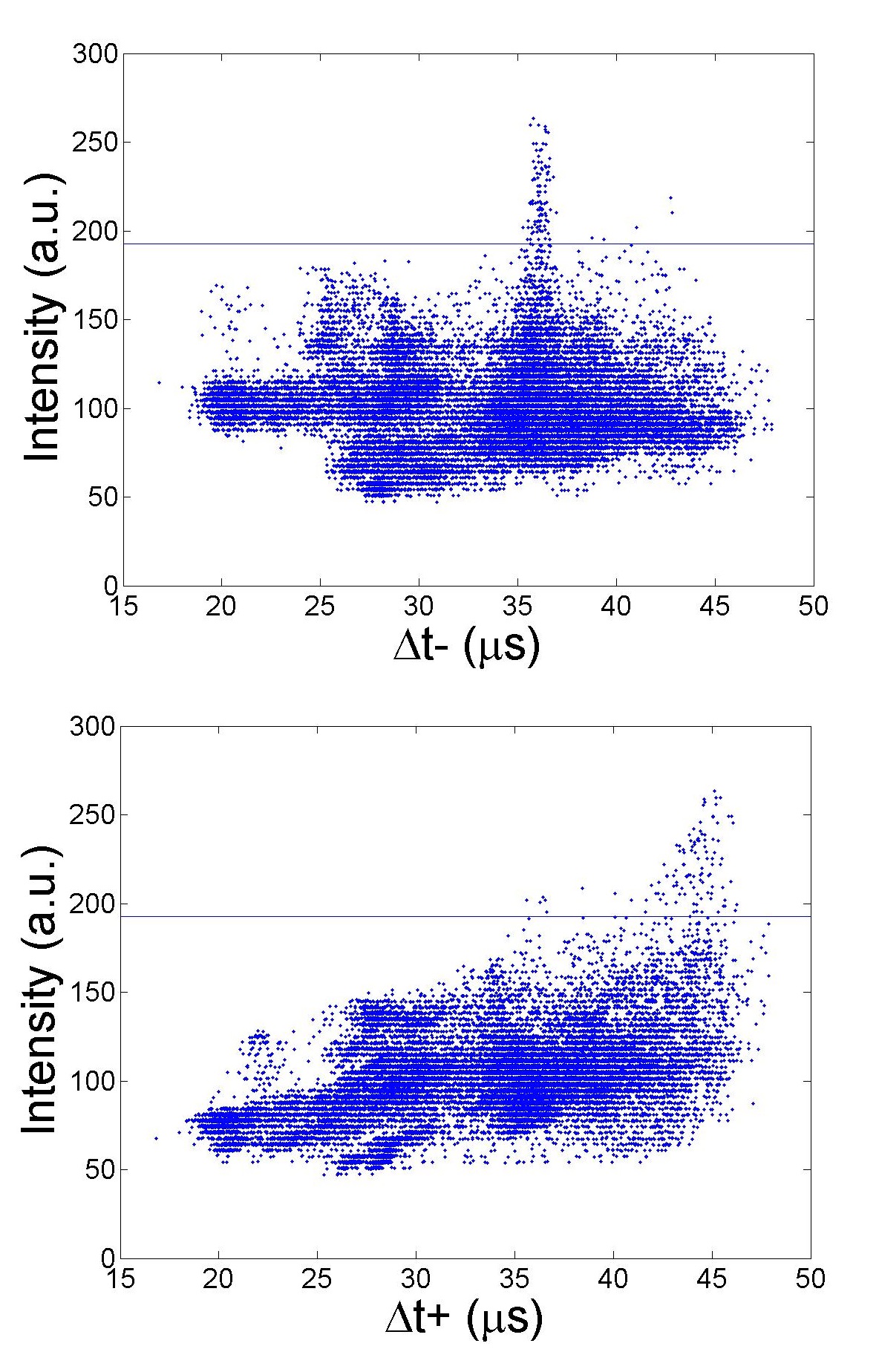}
\caption{\label{fig:retorno} Plots for peak pulse intensity as a function of interpulse time interval for a chaotic time series 
with EEs. Upper (lower) figure shows the peak intensity of each pulse as a function of $\Delta t -$
($\Delta t +$), i.e. the temporal interval between said pulse and the previous (next) one. The horizontal line indicates the 
rogue wave limit. Peak intensities are expressed in arbitrary units scaled so that an average 
event has an intensity of 100.}
\end{figure}

Figure \ref{fig:retorno} shows, for a typical chaotic 
regime with EEs, plots of the intensity of each pulse against the temporal interval between that pulse and the 
previous one ($\Delta t -$, upper graphic), and between that pulse and the next one ($\Delta t +$, lower graphic). The 
intensity is scaled so that its average value is 100 in 
arbitrary units. Note that both 
plots exhibit a remarkable regularity: the width of the range of values of $\Delta t -$ and $\Delta t +$ associated to extreme 
events is far 
narrower (roughly 2 and 5 $\mu s$ respectively) than that related to, e.g., average pulses (28 and 25 $\mu s$). 
This is evident for the case of $\Delta t -$. This means that: 
i) if one knows that the next pulse is going to be an extreme one, then one is able to predict when it is 
going to occur; ii) once an EE has happened, one can predict the time that it 
takes for the build-up of the next pulse (which is most probably \textit{not} an EE) to appear. If an average pulse is 
considered instead, none of these predictions can be made.

Moreover, the plots provide some hints on the mechanism of generation of EEs. The first one, given 
by the upper plot, is somewhat counterintuitive: in a simplistic approach, one would expect that, the 
higher the peak intensity, the higher the $\Delta t -$, to allow a longer time to accumulate energy from the CW. However, the 
$\Delta t -_{EE}$ (i.e. $\Delta t -$ intervals related to EEs) 
are not particularly long; in fact, they are closer to the average interval than to the maximum $\Delta t -$, since $\Delta t -
_{EE}\simeq$ 36 $\mu s$, while the longest intervals have a value of almost 48 $\mu s$. This suggests that an average pulse does 
not 
totally deplete the energy available in the gain medium, but instead, that it leaves some energy stored in certain regions 
(possibly due to spatial ``hole burning$"$). This stored energy allows a pulse to be an EE in spite of the relatively short time elapsed 
since the last one. Therefore, EEs can be thought as pulses that are more efficient in extracting the energy stored in the gain 
medium. The lower plot shows that the range of values 
of $\Delta t +$ associated to EEs is among the highest of the set, i.e. the time until the next pulse after an 
EE is always among the longest times that can be expected for a particular time series (typically $\Delta t + 
\geqslant$ 40 $\mu s$, while the average $\Delta t +$ is 33 $\mu s$).
This means that EEs efficiently deplete the gain, so that the time until the next pulse is necessarily long, to allow a new 
accumulation of gain. 

\begin{figure}
\includegraphics[scale=0.23]{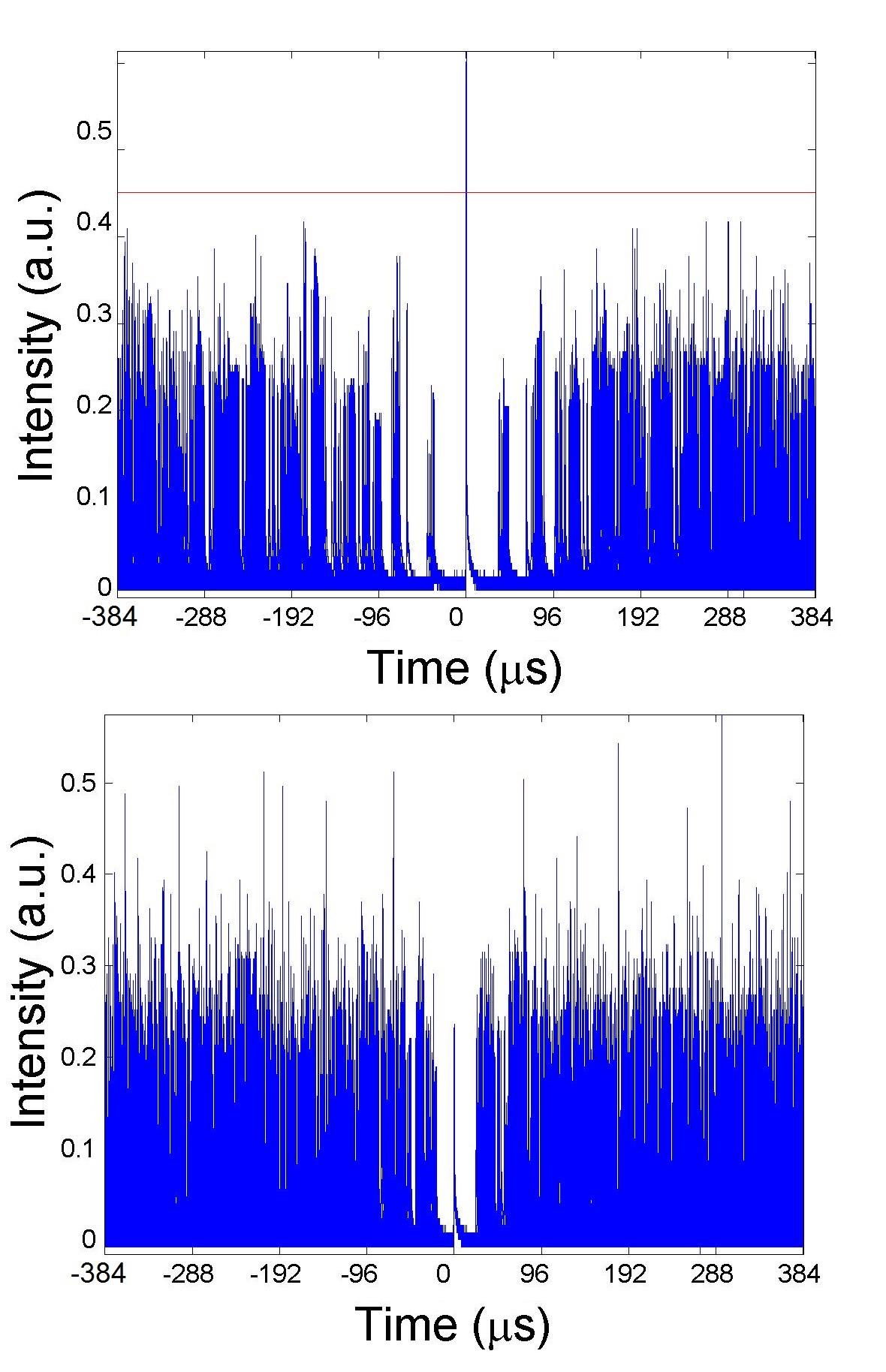}
\caption{\label{fig:sup} Upper plot: superposition of 112 time traces centered at each of the extreme events of a chaotic regime 
of operation (the 
horizontal line shows the rogue wave limit); lower plot: superposition of time traces centered at 112 near-to-average events for 
the same regime of 
operation as in the upper plot.}
\end{figure}

The plots in fig. \ref{fig:retorno} show what happens in the immediate vicinity of a given pulse (i.e. one pulse before or one 
after). It is also convenient to study a wider time scale. 
Fig. \ref{fig:sup} shows, for a time series with 112 EEs, a superposition of time traces 
centered at each of the EEs (upper plot) and, for comparison,                                                                                                                                           
a superposition of the same amount of average pulses (lower plot). Note that all the pulses immediately preceding or following an 
EE (i.e., if the EE is indexed as the Nth event, these pulses are the (N+1)th and (N-1)th events) occur within a fairly narrow 
temporal window, that slowly blurs for farther pulses (the temporal window in which the 
(N+2)th and (N-2)th pulses occur is wider than that of the (N+1)th and the (N-1)th events, and so on). On the contrary, 
for average pulses (lower plot) there is not regular behavior even in the immediate vicinity of the pulse. This suggests that 
the trajectory in phase space corresponding to an EE is confined to a relatively well defined manifold. This result gives 
some hope that EEs can be predict with time enough to control them. 

\subsection{Return intensity maps}\label{sec33}

\begin{figure}
\includegraphics[scale=0.35]{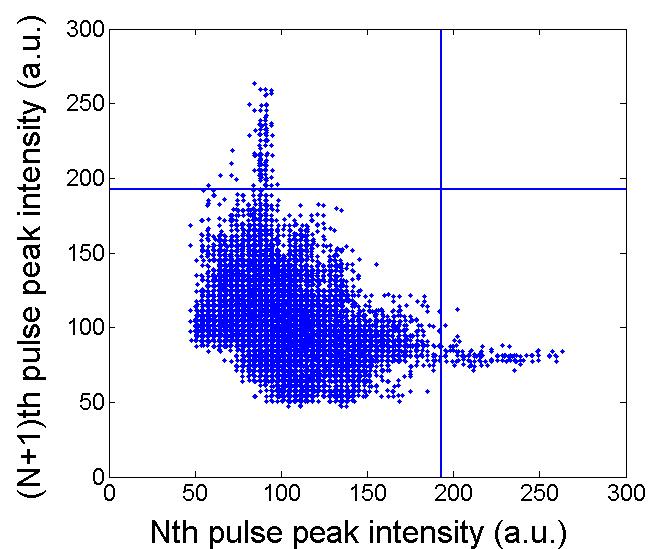}
\caption{\label{fig:mapa} Return intensity map for a chaotic regime of operation with extreme events. The vertical and horizontal line 
indicate the rogue wave 
limit.}
\end{figure}

Return maps, or plots of the intensity of the peak intensity of the (n+1)th pulse 
vs that of the 
nth pulse, provide useful insight. This is shown in fig. \ref{fig:mapa} for the regime 1 of fig. \ref{fig:corresp}. Once again, the EEs show a specific and regular behavior: they appear related to two relatively well localized ``tongues'' emerging from a 
central bunch at a 
value of $\simeq 80$. They are similar to those 
corresponding to the plot of intensity vs interpulse time interval in fig. \ref{fig:retorno}. Therefore, EEs are preceded and 
followed by relatively low intensity pulses (slightly below the average), and the intensity of those pulses is restricted to a 
narrow range of values. This 
means, once again, that the evolution of the peak intensity of the pulses surrounding an EE is quite repetitive. The 
fact that the 
pulse preceding the extreme 
event is close to average supports the picture described in section \ref{sec32}.

In summary, according to the results presented in sections \ref{sec32} and 
\ref{sec33}, the 
EEs are not merely high intensity pulses. They seem to follow a dynamics that is different from that of the average Q-
switch pulse in the temporal series. Furthermore, the dynamics of EEs seems to be more regular than that of the 
average pulses. This enforces the idea that there is a deterministic mechanism behind the formation of EEs.

\subsection{Interferograms}\label{sec34}

\begin{figure}[H]
\includegraphics[scale=0.17]{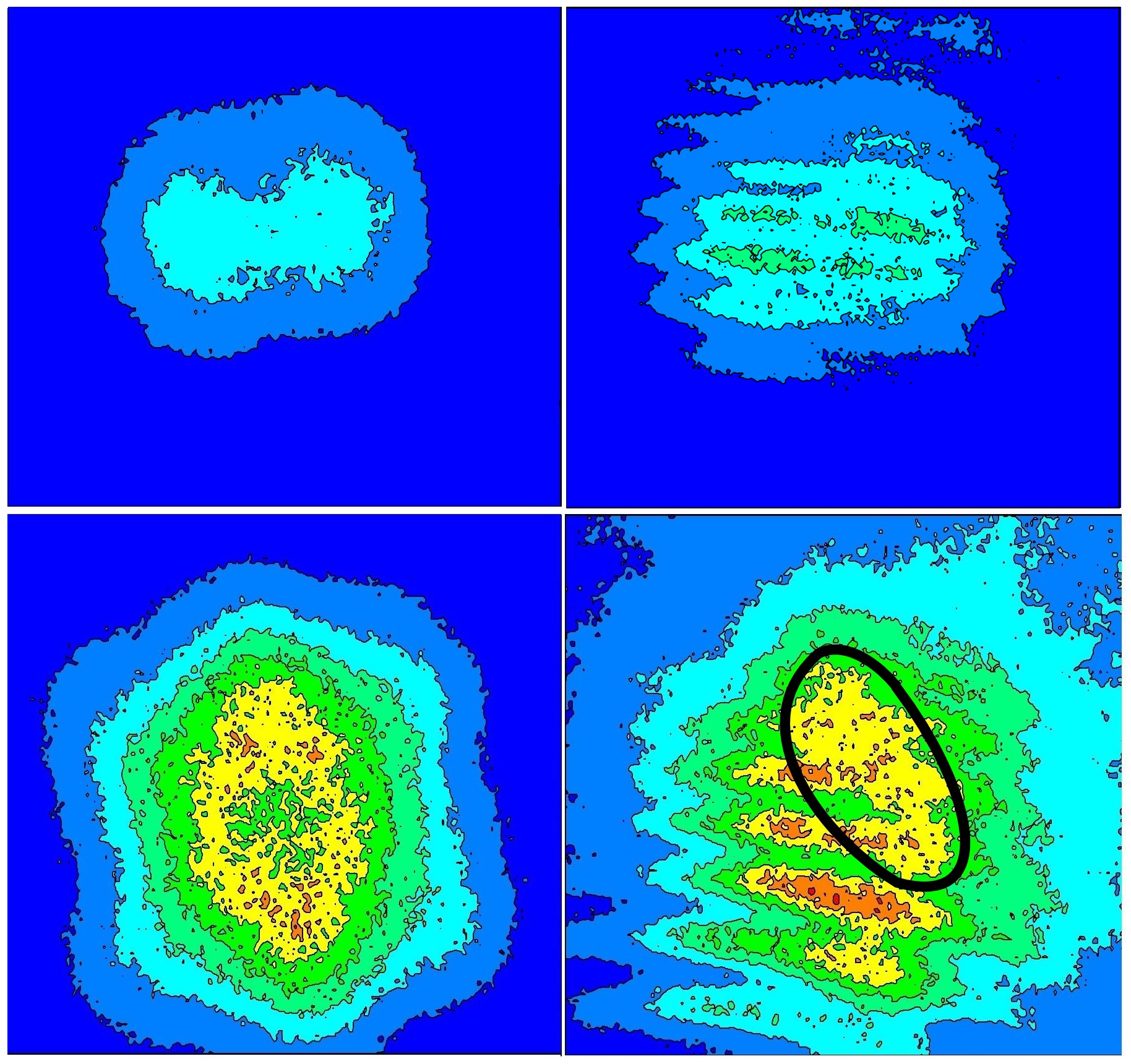}
\caption{\label{fig:interferograma4} Each row corresponds to a different regime; upper row: chaotic regime without 
extreme events; lower row: chaotic regime with extreme events; left column: laser spot; right column: interferogram. The region 
marked with a black line in the lower right figure indicates the region where the interference fringes blur} 
\end{figure}

Fig. \ref{fig:interferograma4} shows the laser spot and heterodyne interferograms, obtained as explained in section \ref{sec2-1} 
for two different 
dynamical regimes: a chaotic regime without rogue waves (upper row), and a 
chaotic regime with 
EEs (lower row). In the former regime, fringe patterns are observed all across the spot. In the latter regime, 
instead, fringes are observed in the region that is expanded on the reference beam and its immediate neighborhood. Outside this 
region, fringes blur and disappear, indicating that the coherence is lost. This result suggests the existence of coherence domains 
and provides further support to the transverse mode interaction hypothesis. 
It is important to remind that, because the transverse pattern varies from pulse to pulse (as it was shown in section 
\ref{sec312}), and due to the long exposure time of the CCD, neither the spot nor the interferogram correspond to a single pulse 
transverse 
pattern, but rather to a superposition of many different ones. We foresee repeating 
this experience using an ultrafast camera in order to distinguish pulse to pulse spots and interferograms. 

\subsection{Dynamical and spatial complexity} \label{sec35}

\begin{figure}
\includegraphics[scale=0.3]{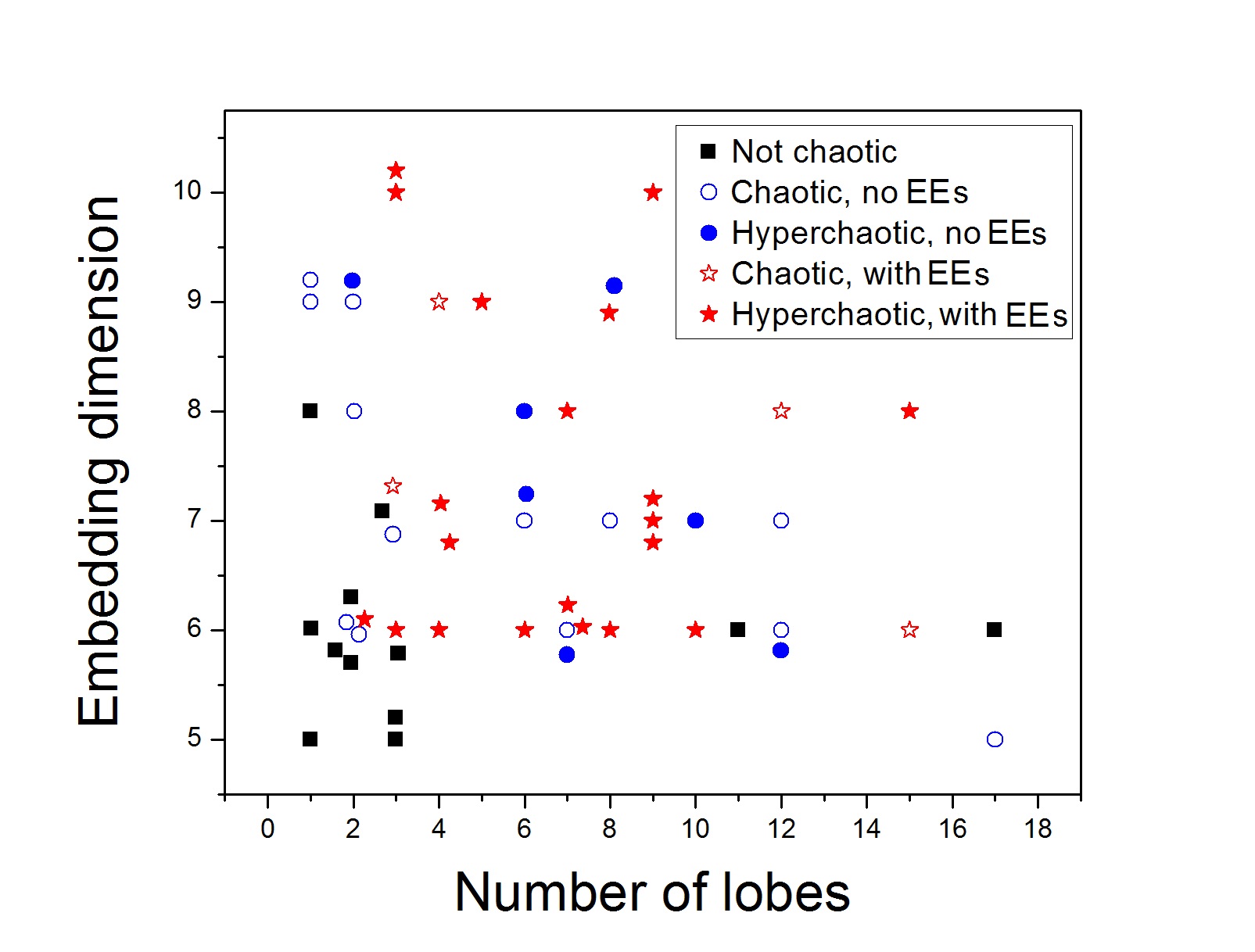}
\caption{\label{fig:diagrama} Representation of all the recorded dynamical regimes in this study having a measurable embedding dimension in terms 
of the number of lobes of their spot (horizontal axis) and their embedding dimension (vertical axis). The color and shape code 
indicates whether the series are not chaotic, i.e. periodic (full black squares), chaotic without extreme events (hollow blue 
circles), hyperchaotic without extreme events (full blue circles), chaotic with extreme events (hollow red stars) or hyperchaotic 
with extreme events.}
\end{figure}

Fig. \ref{fig:diagrama} summarizes the features of all the time series recorded in this study having a measurable embedding 
dimension. They are placed according to their spatial complexity, which is quantified by the number of ``lobes'' in the spot and 
their value of embedding dimension. A color and shape code indicates whether a particular series is non-chaotic, chaotic (one 
positive Lyapunov exponent), hyperchaotic (more than one positive Lyapunov exponent), and whether it displays EEs or not. It can 
be seen, as a general behavior, that chaotic and hyperchaotic regimes have a ``large" dimension of embedding. Here, ``large" 
means higher than the value of 4 predicted by the standard theoretical model based on rate equations for a single mode 
\cite{tang_deterministic_2003}. The time series with EEs tend to display a large number of lobes, too. This 
implies that regimes with EEs are associated with a dynamical behavior which is complex both in space and time. On the 
other hand, chaotic with no EEs and periodic regimes tend to concentrate in the lower left quadrant of the graph, which implies 
they have a lower degree of complexity (lower dimension of embedding and spatially simpler spots). This figure enlarges the 
results presented in \cite{bonazzola_optical_2013-1}.

\section{SUMMARY}\label{sec:conclu}

In this paper, we report a series of observations aimed to guide the way to a (still missing) theoretical explanation of the 
formation of EEs in all-solid-state, self-Q-switched lasers.

In the regime with EEs, the spatial  correlation decays to zero in a distance comparable with the spot size. This is not observed 
in periodic (therefore without EEs) regimes. No decay associated with a narrow peak is observed as it was, instead, in 
other extended spatio-temporal systems. This is consistent with the idea that relatively few
transverse modes are involved in the formation of the EEs. The plots of the intensities recorded in the two detectors, 
or IB vs IA,  show that the transverse patterns change in an irregular way from pulse to pulse, and that the EEs are 
not linked to a specific transverse pattern. 

From the analysis of the interpulse time intervals we see that: i) the values of $\Delta t -_{EE}$ ($\Delta t +_{EE}$) are 
typically contained in a narrow temporal span; ii) $\Delta t -_{EE}$ are next to average, contrarily to what might be expected; 
iii) $\Delta t +_{EE}$, on the other hand, are among the longest interpulse intervals. The result (ii) is consistent with a 
scenario where typical pulses do not totally deplete the accumulated gain in the active medium, but leave some energy stored, so 
that the following pulse does not need a build-up time proportional to its intensity. This energy storage is presumably related 
with spatial hole burning in the active medium. The result (iii) suggests that the deep cause of the EEs is simply 
that they are particularly efficient in depleting the gain, through a mechanism that is to be elucidated.  All these results, 
together with the regular behavior exhibited in fig. \ref{fig:sup}, suggest that there exists a deterministic mechanism in the 
formation of the EEs, and therefore that there is some possibility of predicting and controlling them

The interferograms have shown the existence of domains of coherence compatible with a dynamics ruled by modes interaction. 

Finally, the summarizing diagram Fig. \ref{fig:diagrama} shows that the EEs are prone to arise in chaotic dynamic 
regimes with large dimension of embedding and with a spot with a complex transverse structure. This result also supports the few-
mode interaction hypothesis as the basis of the mechanism of formation of EEs in this type of lasers.

We foresee using an ultrafast camera in order to record series of individual pulse spots (as well as single spot interferograms) 
and identify the ones corresponding to EEs. Another planned course of action is to replace the pump laser diode with a 
VCSEL, which provides a much spatially uniform pump mode, and study its effect on the formation of EEs. Our ultimate goal is the 
construction of a theoretical model able to predict the dynamics of EEs in this system.  

\section*{ACKNOWLEDGEMENTS}

This work was supported by the grant FA9550-13-1-0120, ``Nonlinear dynamics of self-pulsing all-solid-state laser'' of the AFOSR 
(USA), the contract PIP2011-077 ``Desarrollo de l\'aseres s\'olidos bombeados por diodos y de algunas de sus aplicaciones'' of the 
CONICET (Argentina), and the project OPTIROC of the ANR (France).

\bibliography{carlosbib}

\end{document}